\documentclass[12pt]{article}

\usepackage{times}
\usepackage{epsfig,amsmath}
\usepackage{subfigure}
\usepackage{graphicx}
\usepackage{color}
\usepackage{epstopdf}

\newenvironment{sciabstract}{%
\begin{quote} \bf}
{\end{quote}}

\newcounter{lastnote}

\title{Direct Observation of Quantum Percolation Dynamics}
\author
{Zhen Feng,$^{1,2}$ Bing-Hong Wu,$^{1}$ Hao Tang,$^{1,2}$ Lu-Feng Qiao,$^{1,2}$ Xiao-Wei Wang,$^{1,2}$\\ 
	Xiao-Yun Xu,$^{1,2}$ Zhi-Qiang Jiao,$^{1,2}$ Jun Gao,$^{1,2}$ and Xian-Min Jin$^{1,2,\dagger}$\\
	\\
\normalsize{$^1$ Center for Integrated Quantum Information Technologies (IQIT), School of Physics}\\
\normalsize{and Astronomy and State Key Laboratory of Advanced Optical Communication}\\
\normalsize{Systems and Networks, Shanghai Jiao Tong University, Shanghai 200240, China}\\
\normalsize{$^2$CAS Center for Excellence and Synergetic Innovation Center in Quantum Information}\\ 
	\normalsize{and Quantum Physics, University of Science and Technology of China, Hefei 230026, China}\\
\normalsize{$^\dagger$E-mail: xianmin.jin@sjtu.edu.cn}\\
}
\begin{document}
\baselineskip24pt

\maketitle

\begin{sciabstract}
Percolation, describing critical behaviors of phase transition in a geometrical context, prompts wide investigations in natural and social networks as a fundamental model. The introduction of quantum-intrinsic interference and tunneling brings percolation into quantum regime with more fascinating phenomena and unique features, which, however, hasn't been experimentally explored yet. Here we present an experimental demonstration of quantum transport in hexagonal percolation lattices by successfully mapping such large-scale porous structures into a photonic chip using femtosecond laser direct writing techniques. A quantum percolation threshold of 80\% is observed in the prototyped laser-written lattices with up to 1,600 waveguides, which is significantly larger than the classical counterpart of 63\%. We also investigate the spatial confinement by localization parameters and exhibit the transition from ballistic to diffusive propagation with the decrease of the occupation probability. Direct observation of quantum percolation may deepen the understanding of the relation among materials, quantum transport, geometric quenching, disorder and localization, and inspire applications for quantum technologies.
\end{sciabstract}

\noindent Percolation describes an abrupt transition from a disconnected state to connectivity, unveiling the simplest and most fundamental phenomenon in phase transition in statistical physics \cite{Saberi2015}. The liquid flow in porous media was first studied as a percolation process in $1957$ and a most important concept of percolation threshold was proposed to define the critical void fraction where permeation first occurs \cite{BroadbentHammersley1957}. Percolation theory governs a simple rule that structure lattices, with vacant (unavailable) or occupied (available) sites, follow a binomial probability distribution. The seemingly unrelated geometric structure determines a long-ranged correlation in the system at the vicinity of threshold, where rich phenomena occur \cite{Saberi2015}. This facilitates broad and deep understandings in wide ranges of areas: spanning from natural science (e.g. the fractal coastlines \cite{Sapoval2004,Saberi2013}, core formation mechanism \cite{Mann2008}, turbulence \cite{Bernard2006}) to applied science (e.g. conducting materials \cite{Vigolo2005, Grimaldi2006}, colloids \cite{Gnan2014}, magnetic models \cite{Fortuin1972}), and even to social science (e.g. epidemic spreading \cite{Pastor2015}).

\begin{figure*}[htbp]
	\centering
	\includegraphics[width=1\textwidth]{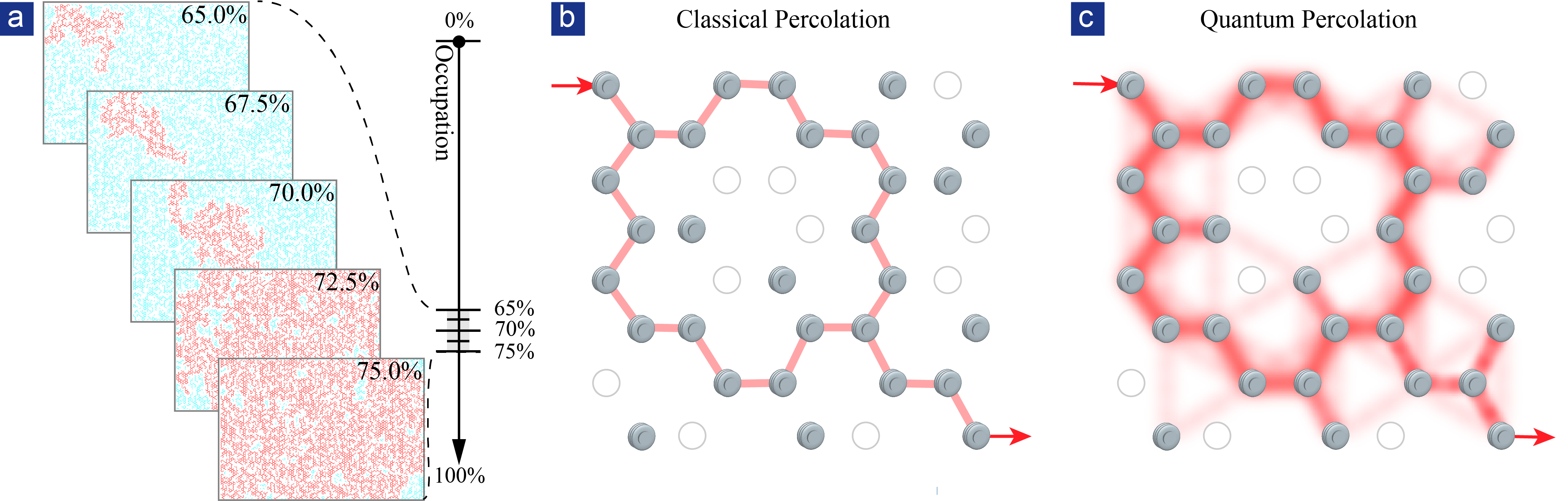}
	\caption{\textbf{Classical and quantum percolation.} \textbf{(a)} Classical percolation is simulated in a hexagonal lattice and the largest cluster is denoted in red. \textbf{(b-c)} A classical particle (no coherence) is injected into the percolation lattice where gray sites are occupied and white ones are empty. A percolation completes when the particle penetrates the lattice and arrives at the other end. In classical percolation \textbf{(b)}, an infinite path spanning across the graph means a complete percolation, while in quantum percolation \textbf{(c)}, photon can interfere with itself through all possible paths, and can access to classically forbidden paths by tunelling.}
	\label{fig:figure1}
\end{figure*}

Transport medium in percolation theory is modeled by a regular lattice, and its relevant entities determine whether the type is site or bond percolation. The former is more general, since every bond model is equivalent to a site one on a different graph, but not vice versa. We therefore are able to investigate a general quantum percolation properties in site percolation lattices. Each site is occupied independently and randomly with a probability $P$ and empty with $1-P$. A cluster is identified to connect the nearest-neighbour occupied sites, and the largest one is denoted in red in Fig.\ref{fig:figure1}(a), where the cluster leads to an explosion when the occupation probability reaches $70\%$ in a simulated $256,000$-sited hexagonal lattice.

The schematic of a small-scale site percolation is shown in Fig.\ref{fig:figure1}(b). A classical picture of penetration means that a cluster spans across from the top left to the bottom right \cite{Essam1980}. As its counterpart, quantum percolation is endowed unique features by inherent quantum inference and tunneling of single or multi particles. Fig.\ref{fig:figure1}(c) illustrates a sophisticated process that a quantum particle injected from the top left interferes with all possible paths, and tunnels into the classically forbidden paths. The percolation Hamiltonian can be represented by
\begin{equation}
H=\sum\limits_{[i,j]} t_{ij} a_i^\dagger a_j+c.c.
\end{equation}
where we define a new notion $[i,j]$ to include two type of relations between the present site $i$ and the nearest-neighbour or the next-to-nearest-neighbour occupied site $j$, and their coupling coefficient is $t_{ij}$.

In regular lattices without empty sites, a quantum walker can achieve a remarkable speedup over its classical counterpart in virtue of the quantum interference and superposition, such as in one dimensional ordered chain \cite{Perets2008}, square lattice \cite{Tang2018b} and glued tree \cite{Tang2018a}. Disorder, however, can induce a transition from quantum ballistic to classical diffusive transport in a quantum walk, and therefore suppress the expansion of the quantum mechanical wavepackets exponentially \cite{Schwartz2007,Schreiber2011}. 

Anderson model randomly imposes a small amount of disorder $\epsilon$ on all sites, and the interference between multiple-scattering paths results in localization. The Hamiltonian is given by $H=\sum\limits_i \epsilon a_i^\dagger a_i+\sum\limits_i\sum\limits_j t_{ij}a_i^\dagger a_j$ \cite{Anderson1958}. Quantum percolation with quenched geometric disorder is essentially different from Anderson localization. Empty sites act as high-potential barriers and form many boundaries in the graph. Superposition of quantum particles in different paths also leads to localization, which suppresses the wavepacket spreading. Inversely, tunneling allows quantum particles pass through the potential barriers that classical particles cannot surmount (e.g. paths between the next-nearest occupied sites) and this enhances the penetration. 

\begin{figure*}[htbp]
	\centering
	\includegraphics[width=1\textwidth]{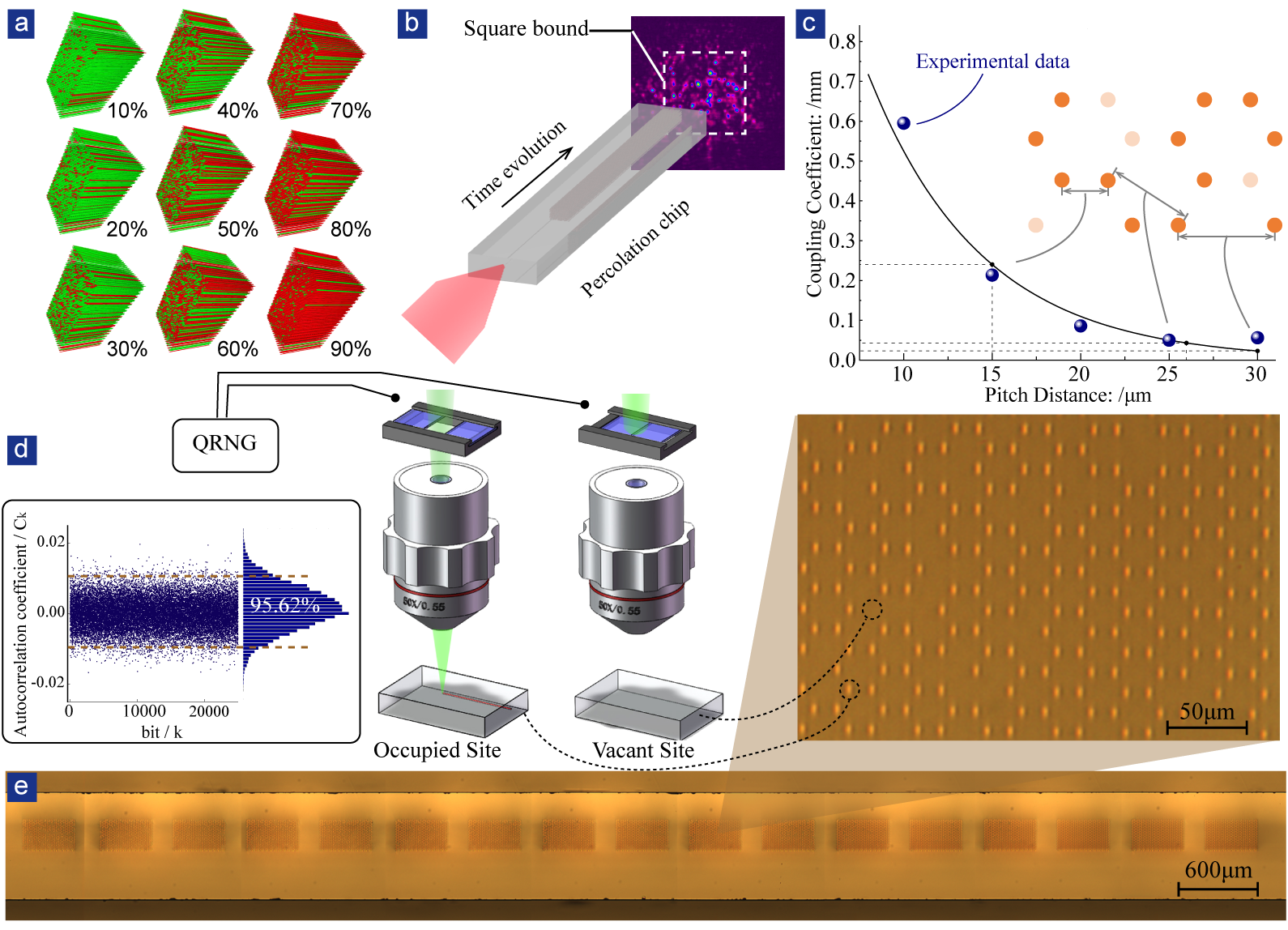}
	\caption{\textbf{Constructing large-scale percolation lattices on a photonic chip (a)} Simulated lattices with different occupation probabilities. Red and green denote occupied and vacant respectively. \textbf{(b)} Schematic of the quantum percolation experiment. Photon injected into a central waveguide evolves in the percolation waveguide structure. The transverse distribution at the output is collected by an optical beam profiler.\textbf{(c)} The relation between pitch distances and coupling coefficients. The blue dots are experimental data. \textbf{(d)} Programmable fabrication of percolation lattice. A shutter is linked to a quantum random number generator (QRNG) to control the writing laser on and off. The independence and randomness of occupied distribution are ensured when the random number can well pass the autocorrelation test. \textbf{(e)} A photonic chip with many $85\%$-occupied percolation lattices. The inset shows the zoomed picture for part of a lattice.}
	\label{fig:figure2}
\end{figure*}

The competition of these quantum effects associated with graph geometries endows quantum percolation \cite{Chakrabarti2009} important roles in understanding quantum transport in condensed matter and material physics, as well as biological and chemical systems \cite{Engel2007,Mohseni2008,Plenio2008}. Quantum percolation model has been comprehensively investigated in theory \cite{Chakrabarti2009,Chandrashekar2014,Thomas2017,Qi2019,Pant2019}. However, it is still experimentally challenging to realize a large-scale quantum network and the required random quenches with a controllable fashion (Fig.\ref{fig:figure2}(a)), whereas a proper analogue and comparison to classical percolation remain to be built.

We manage to demonstrate an experimental quantum percolation and directly observe quantum transport transition in hexagonal percolation lattices by successfully mapping such large-scale porous structures into a photonic chip. The simulated lattices with different occupation probabilities (Fig.\ref{fig:figure2}(a)), containing $1,600$ static sites for each, are successfully mapped into a photonic chip in a controllable and programmable fashion. Schematic of experimental setup is shown in Fig.\ref{fig:figure2}(b). Photon immune to environmental interaction is an ideal quantum particle, and its time evolution in two dimension can be mapped along the propagation distance with an invariant structure, where the coupling coefficients here are plotted in Fig.\ref{fig:figure2}(c). Time evolution in the $x$-$y$ (transverse) plane is represented by transverse distribution along the propagation axis and captured by the optical beam profiler. We realize the three-dimensional prototyping and sophisticated parameter control by using femtosecond laser direct writing \cite{Davis1996,Nolte2003,Crespi2013,Chaboyer2015}. As is shown in Fig.\ref{fig:figure2}(d), a quantum random number generator is imbedded to control the laser whether to process the waveguide or to leave it vacant. By this method, we fabricate a series of $40\times 40$-waveguide percolation structures (Fig.\ref{fig:figure2}(e)) and inject our $810nm$ photons into a central waveguide to inspect quantum percolation for each lattice on chip. 

In order to provide a classical counterpart for fair comparison, we construct a classical percolation model in the same lattice structure. A modest amount of viscous liquid is pumped from a central pipe into the hexagonal percolation pipe network where pipes connect every two nearest occupied sites, assuming that it flows at a constant velocity in the network. The detailed mathematic derivation and data analysis of classical peroclation model is provided in Supplementary Materials.

To explore the confinement property in each occupation probability, we introduce Inverse Participation Ratio (IPR), which takes almost full information about the Hilbert space and is described mathematically by 
\begin{equation}
IPR=\frac{\sum_i |\psi_i|^4}{(\sum_i |\psi_i|^2)^2}
\end{equation}
where $|\psi_i|^2$ represents the intensity distribution of the transverse section \cite{Schwartz2007}. A state localized in one site is the maximum $IPR=1$, and it spreading over $N$ sites approximately has the minimum $IPR=1/N$. $IPR$ has the units of inverse area, thus an average effective width $\omega_{eff} = \langle IPR\rangle^{-1/2}$ represents the statistical average in localization length of multiple experiments with the same occupation probability.

\begin{figure*}[htbp]
	\centering
	\includegraphics[width=1\textwidth]{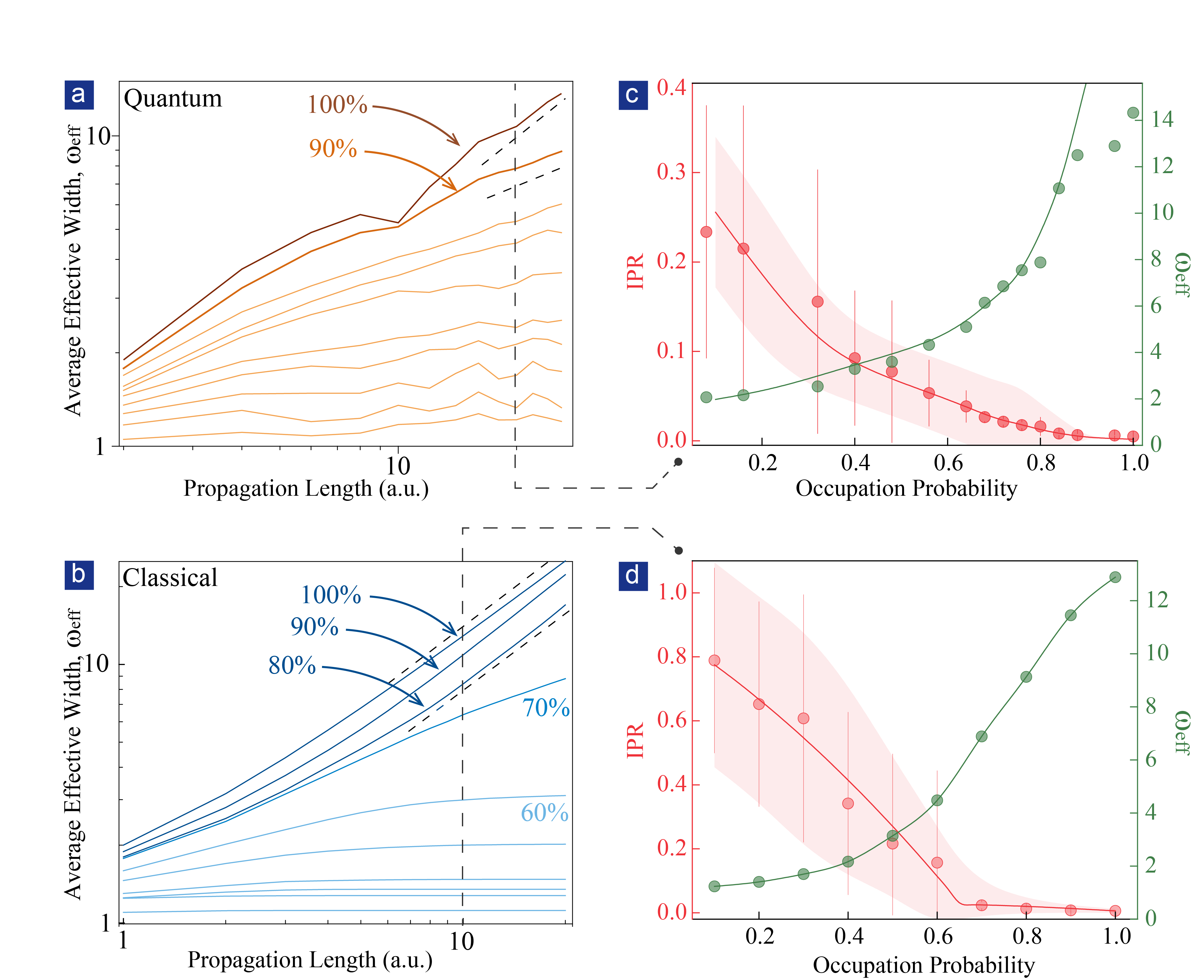}
	\caption{\textbf{Evolution features in quantum and classical percolation. (a-b) Time evolution of average effective width. (a)} Different curves represent different $P$ values and the slope $\nu=1, \nu=\frac{1}{2}$ show a ballistic and a diffusive transport respectively. It is localized when $0<\nu<\frac{1}{2}$. \textbf{(b)} The slope remains at $1$ when $P>80\%$ and drops rapidly under $70\%$, showing a great accordance with the predicted threshold at $69.62\%$ and also with $63\%$ in our model. In the propagation length of $20~mm$, we retrieve the different distributions of inverse participation ratio (IPR) and effective widths $\omega_{eff}$ in the \textbf(c) quantum and \textbf(d) classical percolation. All the points in \textbf{(c)} are experimental data.}
	\label{fig:figure3}
\end{figure*}

We average $\omega_{eff}$ from tens of transverse intensity distributions as a function of propagation length to investigate the time evolution of quantum and classical percolation. The results are shown in Fig.\ref{fig:figure3}(a-b) with each curve representing a different occupation probability from $10\%$ to $100\%$ spacing $10\%$. It is obvious that average effective widths increase monotonously with propagation length and expand faster with a higher occupation probability. We set the coordinate axis as a double-logarithmic scale to indicate the power-law relation $\omega_{eff}\propto z^{\nu}$, where $\nu$ is the slope of the curve. In quantum percolation model, photons performs a ballistic transport in an all-present hexagonal lattice and asymptotically approaches the upper dashed line $\nu=1$. With some sites substituted by vacancy, the decrease of the slope $\nu$ indicates that the broadening is suppressed in different levels: photons have a diffusive spreading with $\nu=\frac{1}{2}$ for $P=90\%$ and a slow-down evolution gradually approaches to a localization with $\nu<\frac{1}{2}$ when occupation probability decreases. On the other hand, the intercepts with the $Y$ axis are monotonous to the occupation probabilities, which conforms to the principle of statistical physics.

We then fix the propagation length and look into the localization property in each occupation probability. We calculate $IPR$ and $\omega_{eff}$ in the quantum scenario, shown in Fig.\ref{fig:figure3}(c). The $\omega_{eff}$ expands almost exponentially with the increase of the occupation probability. Experimental results denoted by circles match well with the simulated lines. The slight deviation in the high-occupation region is attributed to the fact that photons reach the outer edge due to limited lattice scale. In this figure, $IPR$ is a downward convex function, indicating confinement is not very tight under a high-occupation probability $P$. It is distinctly different from the upward convex function in Anderson localization, indicating a sensitive system to a small disorder \cite{Schwartz2007}. We also observe that $IPR$ varies smoothly with $P$ as theoretically predicted in two-dimensional quantum percolation model \cite{Thomas2016}.

In addition, the measured standard deviation $\Delta IPR$ (noted as error bars) is found at the same order of $\langle IPR\rangle$. This result conforms to the prediction that $\Delta IPR$ is inversely proportional to the occupation probability and the ratio of $\frac{\Delta IPR}{\langle IPR\rangle}$ is close to unity \cite{Mirlin2000}. Thus, a relatively large error bar with inverse proportion to $P$ is expected. Fig.\ref{fig:figure3}(d) shows the confinement properties of classical percolation. We can see that $IPR$ approximately fits a quadratic polynomial to the marked points and drops to the base at about $63\%$ and then keeps flat, which reveals a percolation threshold via the turning point in this model. Similar results in average effective width $\omega_{eff}$ also confirm this finding.

To find quantum percolation threshold, we select a boundary centered at the injection point and establish a criteria that a given portion of the injection intensity percolates outside the bound. In our experiment, we set the square bound with a side length of $16$ unit-pitch out of a $40\times 40$ waveguide lattice as the percolation bound (Fig.\ref{fig:figure2}), and the given portion is set as $10\%$ along the propagation distance of $20mm$ correspondingly. This setting does not change the nature of percolation and only results in a small shift along the lateral axis.

\begin{figure*}[htbp]
	\centering
	\includegraphics[width=1\textwidth]{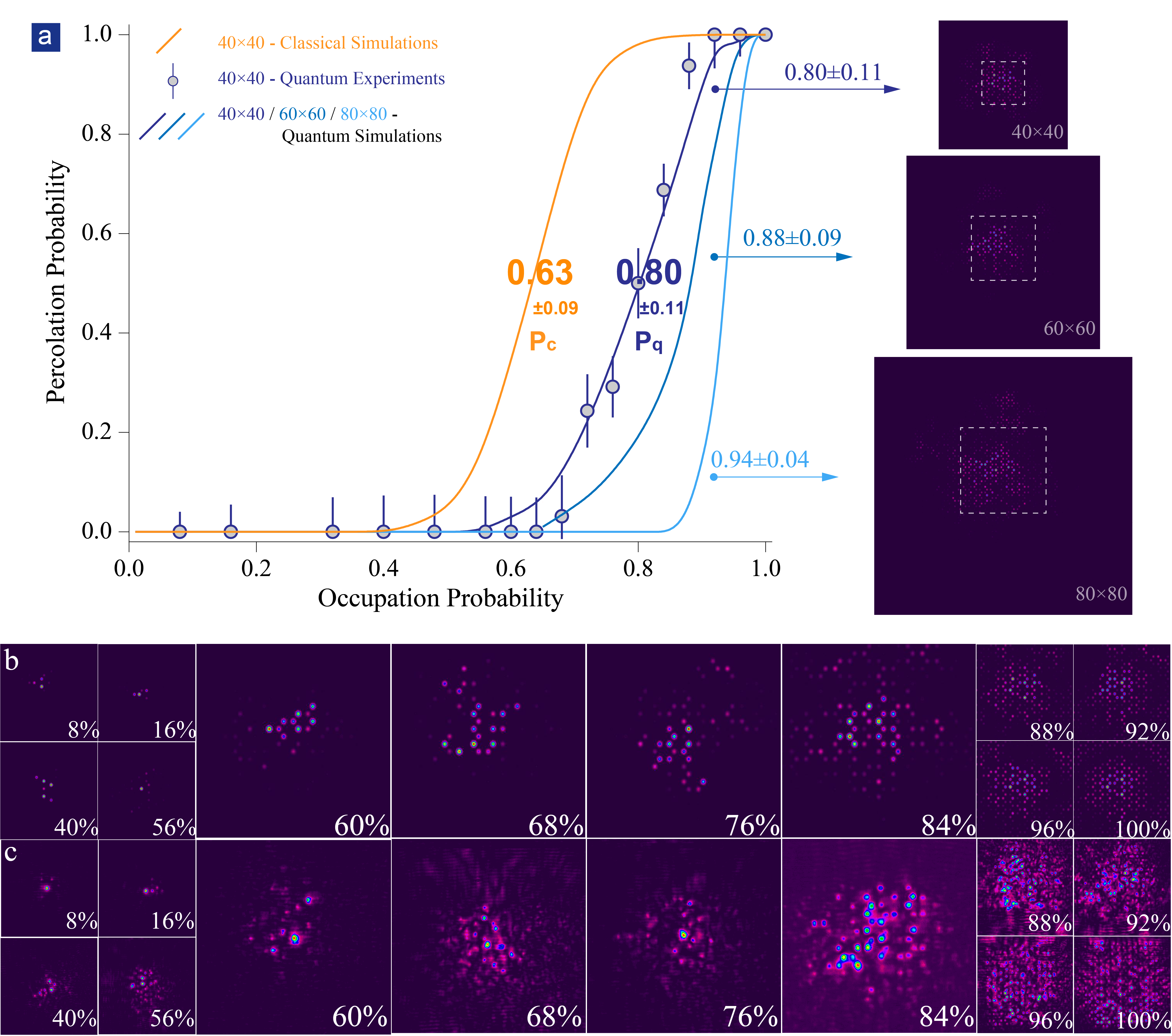}
	\caption{\textbf{Experimental observation of a transition in quantum percolation.(a)} Experimental and theoretical results both demonstrate a sharp quantum percolation transition in the occupation probability range of $80\pm11\%$, i.e. from $69\%$ to $91\%$. Error bars are obtained by error transfer function, see derivation in Supplementary Materials. With the increase of lattice size, the transition turns sharp and the threshold shifts to the right. The cross sections in each scale are attached. Transverse sections for different occupation probabilities are shown in \textbf{(b)}simulations and \textbf{(c)}experiments.}
	\label{fig:figure4}
\end{figure*}

We gather about $40$ individual experimental results for each occupation probability and analyze them one by one. From all the processed data of each value $P$, we count the number of complete penetrations and define this fraction as percolation probability. We generate $100$ results for each occupation probability and form a smooth simulation curve. From statistical and visual results retrieved in a $40\times40$ hexagonal lattice (Fig.\ref{fig:figure4}(a-c)), we can see a clear quantum percolation transition centered at $80\%$, which is much higher than the results of 63\% obtained from the classical percolation model. The transition is quite slow with a broad percolation span, which is attributed to the restricted lattice size. The span length is defined as the experienced range of occupation probability $P$, where the percolation probability rise from $10\%$ to $90\%$ with a fixed size of hexagonal lattices. In our experiment, we have achieved a size up to $40\times 40$, which leads to a span length of $\pm11\%$. We enlarge the lattice size to $60\times 60$ and $80\times 80$ as well as the corresponding scaling of propagation length and the bound size, see detailed parameters in TABLE.\ref{tab1}). The percolation thresholds are found to be pushed to right further and the span lengths become shorter, which well conforms to the theories \cite{Islam2008}.

In conclusion, we present an experimental demonstration of quantum percolation and directly observe quantum transport transition in on-chip large-scale hexagonal lattices. The measured quantum percolation threshold of 80\% in lattices with up to 1,600 waveguides is significantly larger than the classical counterpart of 63\%. We also observe a transition from ballistic to diffusive and then to localized transport with the decrease of occupation probability. The developed platform and capabilities of percolation lattice engineering may open the door to explore rich percolation-involved phenomena for quantum simulation, and inspire applications for analog quantum computing.

Finding exact percolation thresholds has long been attracting enduring explorations for mathematicians because of its fundamental and wide-ranging importance. Early mathematic foundation was established to solve the simplified general Potts model accurately \cite{Kasteleyn1969}, which facilitated many advances in the development of percolation theory, such as scaling relations \cite{Kesten1987,BorgsChayesKestenEtAl2001,BollobasRiordan2006}, the hull \cite{SaleurDuplantier1987} and renormalization group theory \cite{Cardy1996}. Percolation threshold is considered as a fundamental characteristic of percolation theory and its value generally depends on the structure parameters (e.g. lattice parameters and dimensionality) \cite{FisherEssam1961}. Exact thresholds are only known for certain two-dimensional lattices that satisfy particular mathematical transformation condition \cite{SykesEssam1963}. However, a variety of thresholds in many common systems are still missing (e.g. site percolation on hexagonal graphs) \cite{Araujo2014}. Our work provides an alternative approach to solve the threshold problems by building a percolation simulator.


\subsection*{Acknowledgments.}
The authors thank Myungshik Kim, Ian Walmsley and Jian-Wei Pan for helpful discussions. This research was supported by the National Key R\&D Program of China (2019YFA0308700, 2017YFA0303700), the National Natural Science Foundation of China (61734005, 11761141014, 11690033), the Science and Technology Commission of Shanghai Municipality (STCSM) (17JC1400403), and the Shanghai Municipal Education Commission (SMEC) (2017-01-07-00-02- E00049). X.-M.J. acknowledges additional support from a Shanghai talent program.\\


\clearpage

\clearpage

\clearpage

\clearpage

\clearpage
\begin{table*}[htbp]
	\centering
	\caption{\label{tab1}Simulation parameter settings in different-sized lattices}
	\begin{tabular}{c|c|c}
		Lattice size & Propagation length($mm$) &  Square bound size \\
		\hline
		$40\times 40$ & $20$ & $16\times 16$ \\
		$60\times 60$ & $30$ & $24\times 24$\\
		$80\times 80$ & $40$ & $32\times 32$\\
	\end{tabular}
\end{table*}

\newpage
\setcounter{figure}{0}
\renewcommand{\thefigure}{\arabic{figure}}
\renewcommand{\figurename}{Extended Data Figure}

\subsection*{\label{si1}Methods.}
\subsubsection*{\label{si1.1}Fabrication of Percolation Lattices on a Photonic Chip.}
A femtosecond laser ($10W,1026nm$) with $290fs$ pulse duration and $1MHz$ repetition rate is frequency doubled to $513nm$ and directed into an spatial light modulator(SLM) to create burst trains which is focused on a borosilicate substrate($20mm \times 20mm \times 1mm$) with a $50\times$ objective lens with a numerical aperture of $0.55$ . $40\times 40$ straight waveguides are fabricated at a constant velocity of $10mm/s$ for each percolation lattice, i.e. there are $40$ layers in the borosilicate substrate with $40$ waveguides in each layer. Whether a waveguide is written or not, is programmablely controlled by a quantum random number generator (QRNG). The pitch between two nearest-neighbour waveguides is $15\mu m$. In addition, honeycomb lattice spans as large as $400\mu m$ in depth and great efforts have been made to process depth-independent waveguides through power and Spatial Light Modulator (SLM) compensation.


\subsubsection*{\label{si1.3}Statistics of Quantum Percolation Threshod.}
In our experiments, we gather $N_P$ individual experimental results for each occupation probability and count the number $n$ of complete percolations from all these $N_P$ processed data of each value $P$. The fraction of complete percolations $Pr$ is defined by
\begin{equation}
Pr=\frac{n_P}{N_P} \mbox{($P$ is a given occupation probability)}
\end{equation}
In this case, we expand the expression $Pr$ as:
\begin{equation}
Pr=\frac{\sum_i^{N_P} (P_i)}{N_P}
\end{equation}
where $P_i$ is $1$ when the $i$th experiment percolation is accomplished, otherwise is $0$; $\sum_i^{N_P} (P_i)=n$ ($P_i=0$ or $1$). Independent and identically distributed random variables $P_1, P_2, \cdots, P_i, \cdots$ follows Bernoulli distribution with the occupation probability $P$:
\begin{equation}
\Delta P_i=\sqrt{P(1-P)}
\end{equation}
The error bars are derived by the error transfer function:
\begin{equation}
\Delta Pr=\sqrt{\sum_i^{N_P}(\frac{\partial Pr}{\partial P_i})^2 (\Delta P_i)^2}=\frac{\Delta P}{\sqrt{N_P}}=\frac{\sqrt{P(1-P)}}{\sqrt{N_P}}
\end{equation}





\subsubsection*{\label{si4}Classical Percolation Model in Hexagonal Lattices.}
Here we construct a classical percolation model for hexagonal percolation graph. A modest amount of viscous liquid is pumped from a central pipe into the hexagonal percolation pipe network where pipes connect every two nearest occupied sites, assuming that it flows at a constant velocity in the network. It shares the same analysis as our quantum percolation model and can help gain insight into the percolation in quantum regime.

\begin{figure*}[htbp]
	\centering
	\includegraphics[width=0.99\textwidth]{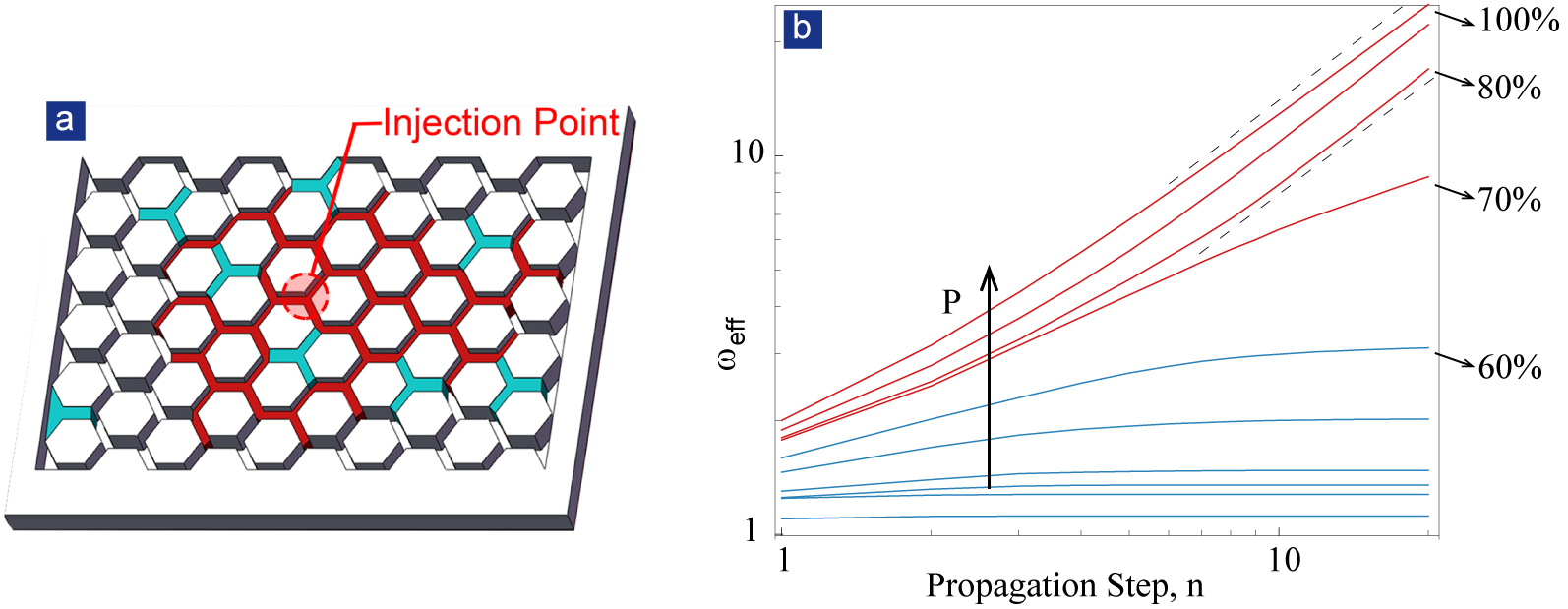}
	\caption{\textbf{Classical Percolation Model in Hexagonal Lattices. (a)} Sketch of hexagonal percolation pipe network. Viscous liquid is pumped into the injection point and evolutes in available paths marked in red. \textbf{(b)} Time evolution in classical percolation model. Three regions are separated by the occupation probability (i) $P > 70\%$, the slopes $\nu$ is going to approach $1$; (ii)$60\% < P < 70\%$, liquid leads to a relatively slow expansion; (iii) $P \leq 60\%$, liquid is trapped in the pipes.}
	\label{fig:figures5}
\end{figure*}

We start with the introduction of inverse participation ratio ($IPR$) to our classical model by
\begin{equation}
IPR=\frac{\int I_{(x,y,L)}^2 dxdy}{[\int I_{(x,y,L)} dxdy]^2}=\frac{\sum^N c^2}{(\sum^N c)^2}
\end{equation}
where $c$, a constant value, represents occupied sites of viscous liquid and $0$ is for vacant site. $N$ denotes the number of occupied sites covered by viscous liquid. Average effective width $\omega_{eff}$ can be derived
\begin{equation}
\omega_{eff}=\langle IPR \rangle^{-\frac{1}{2}}=\sqrt{\frac{(\sum^N c)^2}{\sum^N c^2}}=\sqrt{\frac{N^2 c^2}{N c^2}}=\sqrt N
\end{equation}
In this two-dimensional model, it is apparent that the number of sites $M$ is proportional to $t^2$ in consideration of occupation probability $P$ mentioned in the main text, $t$ is a propagation step. We can obtain the relation:
\begin{equation}
N=P M\propto P t^2
\end{equation}
Then we can get $\omega_{eff}$ as follows:
\begin{equation}
\omega_{eff}=\sqrt N=k\sqrt P t
\end{equation}
where $k$ is a constant factor.
We sketch the relation between propagation step $t$ and average effective width $\omega_{eff}$ in a double-logarithmic scale:
\begin{equation}
\log\omega_{eff}=\log {k\sqrt P t}=\log t+\frac{1}{2}\log{P}+Const.
\end{equation}
In this system, we derive the relation between propagation step and average effective width (Extended Data Fig. \ref{fig:figures5}). We find liquid expands in a ballistic manner, following our theoretical evolution $\omega_{eff}=\sqrt{PM} z$, as long as $P>70\%$. When $P\leq60\%$, a bit lower than the derived classical threshold $63\%$, the liquid will be trapped in the pipes.\\

\clearpage


\end{document}